\documentclass[a4paper,11pt]{article}
\usepackage{epsf,multicol,ifthen}
\usepackage[cp1251]{inputenc}
\usepackage[dvips]{graphicx}
\usepackage{cite}
\usepackage{hyperref}
\newcommand{\bec}{\begin{center}}
\newcommand{\ec}{\end{center}}
\title{ BPS states of Fourfolds as candidates for Kaluza-Klein modes}

\author{T. Obikhod\thanks{E-mail: obikhod@kinr.kiev.ua}\\
\small\emph{Institute for Nuclear Research, National Academy of Science of Ukraine} \\
\small\emph{47, prosp. Nauki, Kiev, 03028, Ukraine}}

\date{\small\today}

\begin{document}

\maketitle

\abstract{Within the framework of the cosmological theory of the Big Bang, F-theory that unifies all four types of fundamental interactions is represented. Among the most exciting predictions of physics beyond the Standard Model is the assumption of the space of extra dimensions that solves the hierarchy problem. With the presence of this extra dimensions are connected the searches for Kaluza-Klein partners of gravitons, gauge bosons and microscopic black holes at the LHC. In the framework of F-theory, Calabi-Yau fourfold is considered  as a space of extra dimensions. We study the duality between the F-theory compactified on the K3-surface and $E_8\times E_8$ heterotic string compactified on the torus, $T^2$. The set of BPS states corresponding to the Calabi-Yau fourfolds, which has either an elliptic curve or a K3-fibration as a layer, is studied in the aspect of correspondence to the KK modes of the M-theory on $R^8\times T^2 \times S^1 / Z_2$. The singularities of the moduli space of the Calabi-Yau fourfold make it possible to observe massive KK modes, the masses of which are obtained from the M-theory of supergravity. The result is of interest for a theoretical understanding of the KK modes, the experimental searches for which are carried out at the LHC. }

\section{Introduction}
\label{sec:intro}

Observational data confirms the correctness of the Big Bang model. The early Universe was filled with radiation and neutrinos with a small number of neutrons and protons. During the expansion, the radiation has cooled down and today its temperature is no more than 3 Kelvin. During cooling, the number of particles over the antiparticles gradually increased. Particle production occurred in an inhomogeneous and anisotropic Universe. After the birth of pairs of particles and antiparticles during the evolution of the Universe, it became increasingly isotropic and homogeneous. In the early Universe, when it was close to the singularity, at high temperature and high radiation density, gravitons could be generated. Their energy was too small to be detected. The second consequence of the early phase of the Universe formation after the Big Bang could be the formation of micro black holes. The space in this case was inhomogeneous and chaotic. According to the chaotic model of Misner's mixed world, there were collapsing regions of all matter in a certain area of space until a black hole formed. At the same time, the masses of such black holes should be relatively small for the appearance of self-destruction which does not violate the global structure of space-time of the Universe. In the next third of a million years, a primary fireball appears, which consisted of primary matter and radiation. Homogeneity and isotropy are confirmed by direct observations of uniform distribution of helium in the Universe.

	Since the theory of the Big Bang is promising among all cosmological theories, one can hope that there will appear a fundamental theory that studies the conditions of the initial singularity. One of the greatest successes of the physics of elementary particles is the combination of electromagnetic and weak interactions into a unified theory of electroweak interaction. The consequence of adding of a third nuclear force is the prediction that during the first $10^{-42}$ seconds after the Big Bang, the temperature was so high that there must exist supermassive particles with mass of $10^{-9}$ grams with equal role of strong, weak and electromagnetic forces. It can be confidently asserted that there are also possible theories with the fourth fundamental force, gravity, which provide for the existence of supermassive particles in an even earlier period after the Big Bang, and in which all four types of fundamental interactions are equally represented.
	
	The theory of strings or M-theory, the theory of supergravity are used, as theories of unified fundamental interactions in one theory of the physics of elementary particles at high energies. However, the problem is that to perform the appropriate experiments, we need the energies that are unreachable at modern accelerators.
	
	Among the most exciting predictions of physics beyond the Standard Model is the assumption of a space of extra dimensions. The advantages of this additional space are that it solves the problem of the hierarchy of interactions, reducing Planck's energy $10^{19}$ GeV to the energy achievable at modern colliders, 10 TeV. With the existence of the space of extra dimensions are associated the searches for such exotic objects at the LHC, as Kaluza-Klein (KK) partners of gravitons, KK partners of gauge bosons and microscopic black holes.

\section{The compactification of F-theory on  fourfold }
	
	We will consider a fourfold as a space of extra dimensions. The choice of this particular space is due to several reasons. First, the choice of the fourfold is due to the "good" holonomy group - either SU(N) or Spin(7), which leads to the conservation of supersymmetry, see Table 1.
\bec
Table 1. {\it Classification of the extra dimensional spaces by holonomy and supersymmetry groups}
\[\begin{tabular}{|c|c|c|c|} \hline
dimension&manifold type&holonomy&SUSY\\ \hline
k&$T^k$&{1}&1\\ \hline
4&$K3$&SU(2)&1/2\\ \hline
6&$CY3$&SU(3)&1/4\\ \hline
7&$G_2$&$G_2$&1/8\\ \hline
8&$Sp(2)$&hyper-K$\ddot{a}$hler&3/16 \\ \hline
8&$CY4$&SU(4)&1/8\\ \hline
8&$Spin(7)$&$Spin(7)$&1/16 \\ \hline
\end{tabular} \]
\ec
	To receive the theory of supergravity in eight dimensions, it is necessary to compactify the 10-dimensional theory of supergravity on the torus $T^2$ \cite{1.}, since the compactification of N = 1 supergravity onto $T^2$ gives N = 1 supergravity in 8-dimensional space.

	Secondly, the compactification of the 12-dimensional theory or the F-theory on fourfold leads to the usual physical four-dimensional space with which the observed physical quantities are associated. The choice of the F-theory as the pretender of the "Theory of Everything", which solves the problem of the hierarchy of interactions, is due to its relation to the $E_8\times E_8$-heterotic string and to the $SO(10)\subset E_8$ symmetry group of the Standard Model. Thus, the compactification of a heterotic string on $T^2$ and the compactification of F-theory on K3-surface gives a dual description of the same set of theories in the eightdimensional space
\bec
\begin{figure}[htbp]
\hspace*{4cm}{\includegraphics[width=.5\textwidth]{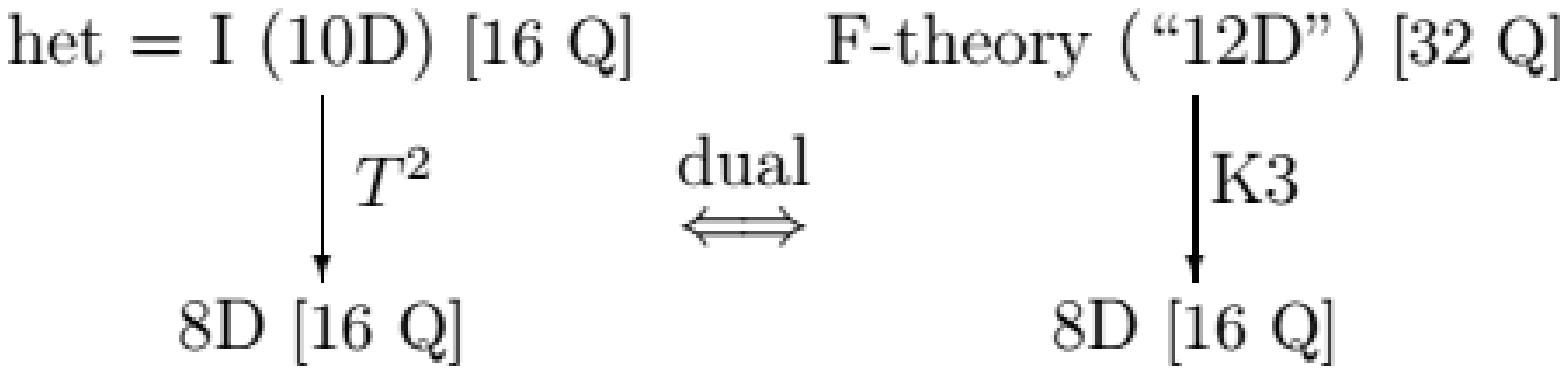}}
\hspace*{3cm} (1)\end{figure}
\ec
The compactification of the F-theory to a supersymmetric theory is associated with an elliptic fibration i.e the vacuum in four dimensions is described by an elliptic fourfold Calabi-Yau. In the future, we'll  work with the Calabi-Yau manifolds, which, as a layer, have either an elliptic curve or a K3-fibration, that is presented by the following form
\bec
\begin{figure}[htbp]
\centering{\includegraphics[width=.25\textwidth]{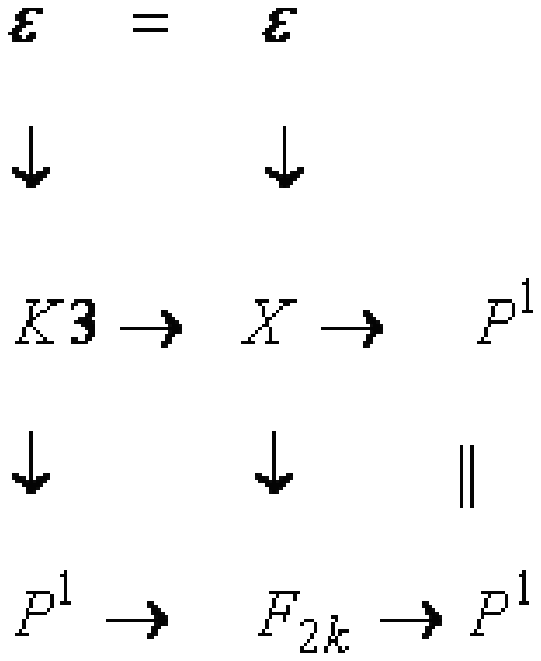}}
\end{figure}
\ec
By its definition, the K3-surface belongs simultaneously to two classes of manifolds that are intensively studied by mathematicians and physicists: the Calabi-Yau manifolds and holomorphic symplectic manifolds. The K3-surface has a 22-dimensional homology group $H_2(K3, Z)$, with a lattice $\Gamma^{3,19}$. Such a space has singularities, the blow-up of which leads to a smooth space. The compactification of an F-theory on a K3-surface is described by an elliptic fibration $X$ over a base $B$
\bec
\begin{figure}[htbp]
\centering{\includegraphics[width=.15\textwidth]{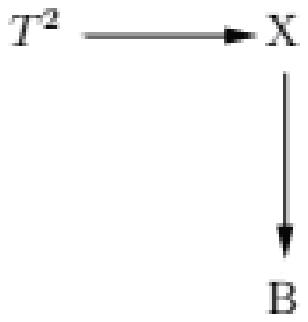}}
\end{figure}
\ec
with a set of elliptic curves (a set of points $(x, y, z)$ in the projective plane  $P^{2,3,1}$ that satisfies the equivalence condition 
\[(x, y, z)\sim (\lambda^{2} x, \lambda^{3} y, \lambda z), \forall\lambda\in C\backslash {0}\] defined at each point of the base, which are parametrized by the Weierstrass form
\[\hspace*{5.3cm}F=-y^2+x^3=f(\omega)x+g(\omega)=0 \hspace*{4cm} (2)\] 
where $f, g$ - are functions of coordinates $\omega$  on the base B. At the singular point, where the discriminant $\Delta=4f^3+27g^2=0$ at the point $(x, y)$  vanishes, the partial derivatives of $F$ with respect to 
$\omega, \frac{\partial F}{\partial\omega}=f^{'}x+g^{'}$  are nonzero, therefore the K3-surface defined by (1) is non-zero at each point $\omega$, although the fiber degenerates at a point $(x, y)$.

	The compactification of the F-theory on a singular K3-surface leads to a vacuum in the eight-dimensional supergravity theory with enhanced symmetry groups, since the presence of a singularity leads to massless gauge bosons classified by such groups. In the M-theory/IIA picture, the gauge boson appears from the membrane/of the D2-brane wrapped around the vanishing cycle, Fig. 1.
\bec
\begin{figure}[htbp]
\centering\includegraphics[width=.55\textwidth]{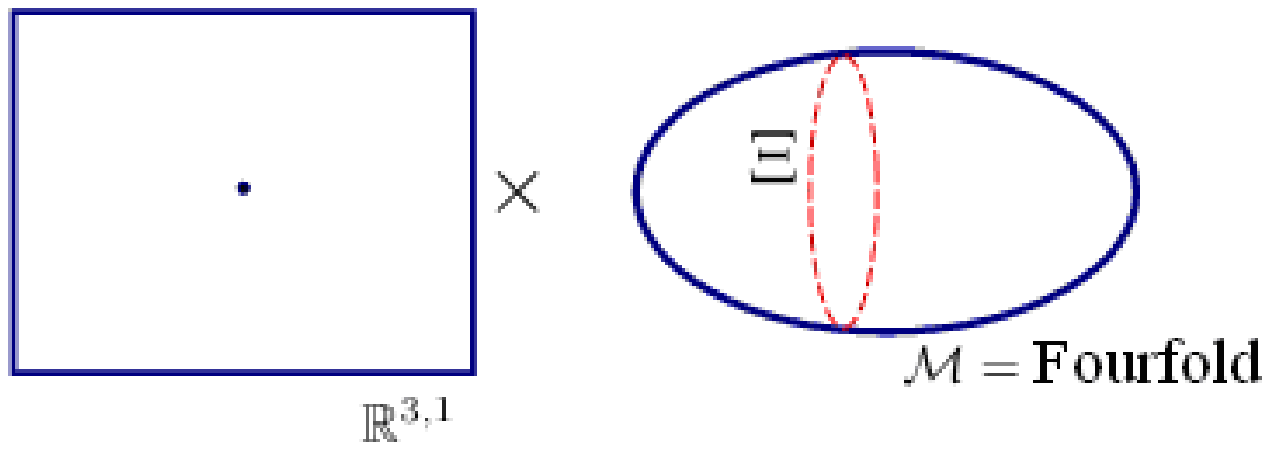}\\
Fig.1 {\small Instantons as euclidean D2-branes, that are point-like in space-time, \\ and wrap the cycle around the fourfold ${\cal{M}}$.}
\end{figure}
\ec
Classification of types of singularities was carried out by Kodaira \cite{2.}, 
with each Dynkin diagram corresponding to each type of singularity. In F-the theory, such Dynkin diagrams characterize a non-Abelian gauge group appearing from the singularities of the K3 surface. A complete set of singularities classified by Kodaira  is presented in Table 2
\bec
Table 2 {\it Table of singularity types for elliptic surfaces
and associated nonabelian symmetry groups} \vspace*{3mm}\\
\begin{tabular}{|c|c|c|c|c|} \hline
ord$(f)$&ord$(g)$&ord$(\Delta$)&$singularity$&nonab. symmetry \\ \hline
$\geq 0$&$\geq 0$&0&none&none \\
0&0&n&$A_{n-1}$&SU(n) \\
$\geq 1$&$\geq 0$&0&none&none \\
$\geq 0$&1&2&none&none \\
1&$\geq 2$&3&$A_1$&SU(2) \\
$\geq 2$&2&4&$A_2$&SU(3) \\
2&3&n+6&$D_{n+4}$&SO(8+2n) \\
$\geq 2$&$\geq 3$&6&$D_4$&SO(8) \\
$\geq 3$&4&8&$E_6$&$E_6$ \\
3&$\geq 5$&9&$E_7$&$E_7$ \\
$\geq 4$&5&10&$E_8$&$E_8$ \\ \hline
\end{tabular}
\end{center}

\section{The duality between the heterotic string and F-theory}

The duality between the heterotic string compactified on the torus $T^2$ and the F-theory compactified on K3, was studied for the first time in \cite{3.}. It leads not only to the same low-energy theory of supergravity, but also to the same physical theory at the nonperturbative level, confirming the existence of a dual symmetry between two classes of string vacuum. And although there are still gaps at the mathematical level in the proofs of the duality between the matter content and the symmetries, suggesting the presence of a deeper geometric structure of these theories, the existence of the duality is beyond doubt.
	Considering the duality between string theory and F-theory, it is necessary to emphasize that light string states correspond to D-branes wrapping around the vanishing cycles. Since, in accordance with (1), there is a duality between the heterotic string and the F-theory, and in the framework of the M-theory all string theories are dual, Fig. 2, then it is reasonable to conclude about the existence of dualities between all the remaining string theories and the F-theory with the corresponding compactification.

For the compactification of the F-theory to the eightdimensional vacuum, it is important to emphasize that Calabi-Yau fourfold is the space of the elliptic fibration defining the F-theory \cite{4.}. This condition was observed in our calculations \cite{5.}, for example, for the Donagi-Grassi-Witten model \cite{4.}, where an elliptic fibration over the base $P^1\times S$  with $S$ - del Pezzo's surface was taken as the fourfold. It is interesting to consider the spectrum of BPS states on the fourfold after the compactification of the F-theory. This issue was studied earlier in \cite{5.} at the level of computations using the INSTANTON computer program, but the theoretical interpretation of this issue remains open. In \cite{6.}, duality with a heterotic string was used for studying of the number of BPS states after the compactification of the F-theory on the Calabi-Yau threefolds. The study was began with $E_8\times E_8$ heterotic strings, which then was compactified to a circle up to 9 dimensions or on the K3-surface. It must be emphasized that during compactification, the $E_8$ group goes over to the $E_d$ group, $E_8\rightarrow E_d \times U(1)^{8-d}$ and the question of the interpretation of new $E_d$ BPS states, which are $E_8$ states deformed on a circle, appears. Similarly, the $E_d$ instantons of a heterotic string compactified on K3 are the dual to instantons of the Calabi-Yau threefolds, where the del Pezzo surface contracts to zero size. For example, among the BPS states there is a string that rotates around the circle, carries one unit of the momentum ($(p_L, p_R)$ mean the left-handed and right-handed momentum of the string, $(p_L, p_R)=\frac{1}{\sqrt{T}}\Biggl(\frac{n}{2R}-mRT, \frac{n}{2R}+mRT\Biggr)$  , where $(n, m)$  are the momentum and the rotation number of the string around the circle (singularity), $T$ is the tension of the string) and becomes massless. At this point there are 252 such states. Since, during our calculations, in \cite{5.} there are repeating sequences of instanton BPS states for Calabi-Yau foulfolds, we can conclude about the equivalence of the qualitative results of the compactification of F-theory on threefolds and fourfolds. However, the appearance of new BPS states for the fourfolds \cite{5.}, signals about the presence of a more complex fibration of Calabi-Yau fourfold, in contrast to the threefold. More complex fibration of Calabi-Yau fourfold leads to new instanton numbers.
\bec
\begin{figure}[htbp]
\centering\includegraphics[width=.35\textwidth]{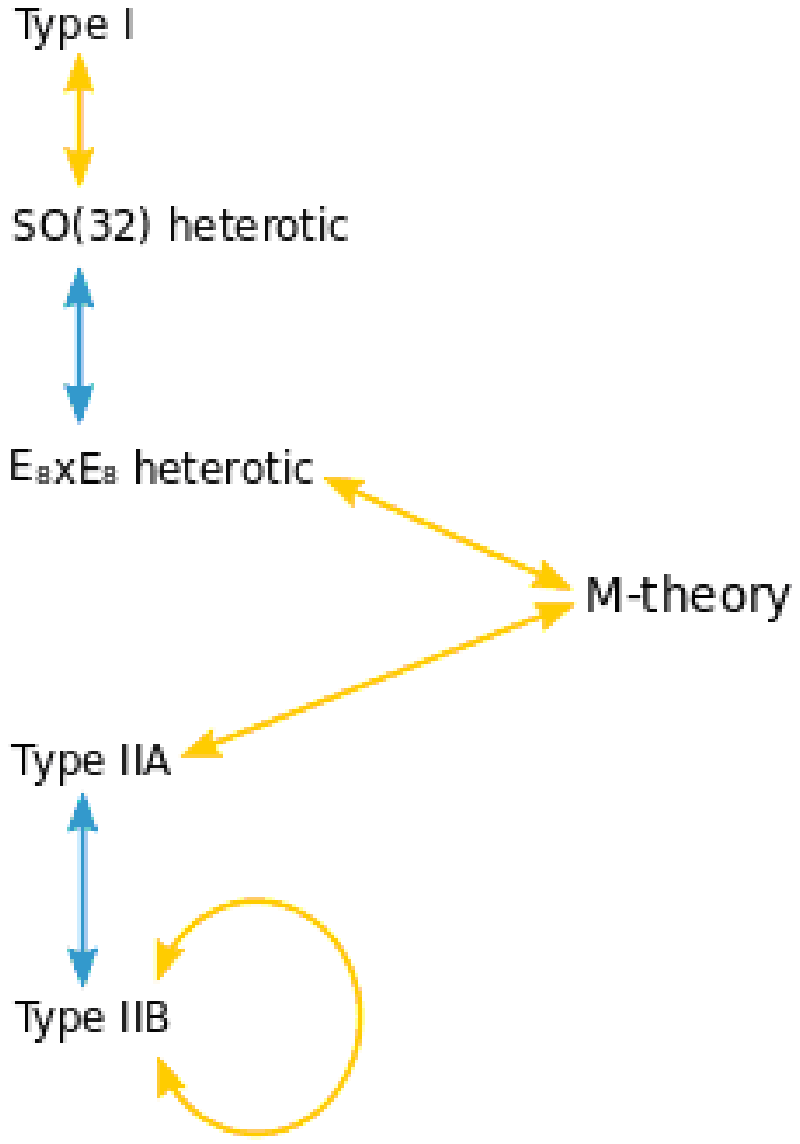}\\
Fig.2 {\small Diagram of dualities between string theories. \\ The yellow arrows are S-duality, the blue ones are T-duality.}
\end{figure}
\ec
\section{Kaluza-Klein states for K3-compactification of the F-theory}
As is known from \cite{7.}, 
\[E_8\times E_8 \rightarrow 
\mbox{M-theory \ on \ } R^{10} \times  S^1/Z_2,\] 
\[E_8\times E_8 \mbox{ \ on \ } R^9\times S^1 \rightarrow \mbox{M-theory  \ on \ } R^9\times 
S^1\times S^1/Z_2   \]
Therefore, 
\[E_8\times E_8 \mbox{\ on } R^8 \times (S^1\times S^1=T^2)\rightarrow 
\mbox{M-theory on\ } R^8\times (S^1\times S^1=T^2) \times S^1/Z_2 .\]
Since we interpreted the heterotic $E_8\times E_8$ string as an M-theory on 
$R^9\times S^1\times S^1/Z_2$, so the M theory on $R^8 \times S^1\times S^1/Z_2$ should be an $E_8\times E_8$ theory on $R^8\times S^1$.
We consider F-theory  compactification on fourfolds, that are obtained from fibering Calabi-Yau threefolds over two-sfere or exceptional divisor, $D$.  As Calabi-Yau, in accordance with the above, has K3-fibration in addition to elliptic fibration, F-theory on K3 is dual to heterotic string on $T^2$ and to the M-theory on $T^2 \times S^1/Z_2$. The K3-compactification of the F-theory leads to singularities with definite two-cycles of the K3-surface that collapse to zero. It leads to massless BPS states. After the K3-compactification, the points of the moduli space representing the conformal field theory and the points representing the theory with enhanced symmetries do not coincide, that is, the dual theory of the heterotic string does not describe the complete string dynamics \cite{8.}. Since we further compactify the M-theory on a circle $S^1$ of radius $R$ to nine dimensions: $R^8\times  T^2 \times S^1/Z_2 \rightarrow E_8\times E_8 \mbox{\ on \ } 
R^8\times T^2 \rightarrow$ F-theory on K3, then we have the equivalence of M-theory with a Calabi-Yau manifold with a K$\ddot{a}$hler class of an elliptic curve $k_E \sim \frac{1}{R}$. In the limit $R\rightarrow \infty$, we obtain the compactification of the F-theory in the eightdimensional space. Particularly interesting is the case of the Calabi-Yau fourfold with $k_D + k_E = 0$, where $k_D$ is the K$\ddot{a}$hler class of the exceptional divisor $D$. At this point, 252 states become massless. According to \cite{6.}, massless 252 states are only the tip of the iceberg, since there is an infinite sequence of massless states 252, -9252, 848628,.... A more detailed study of BPS states at the transition point shows the possibility of supersymmetry breaking and the possibility of the appearance of an infinite number of BPSs-states with mass. 

	From \cite{7.} are known the states in the M-theory on $R^9\times 
	S^1\times S^1/Z_2$. The natural set of states is given by KK states that have masses
\[M^2=\frac{l^2}{R^2_{10}}+\frac{m^2}{R^2_{11}}+n^2R^2_{10}R^2_{11}\]
for certain values of conserved quantum numbers $m$ and $n$ and approximately conserved number l. The masses of KK states in the $E_8\times E_8$ heterotic theory  are given by
\[M^2_{E_8}=\frac{m^2}{R^2_{E_8}}+n^2R^2_{E_8},\]
where the membrane wraping modes are the winding modes of the heterotic string. 
After the compactification onto the torus of the heterotic string, $E_8\times E_8 \mbox{\ on \ } R^8\times T^2 \rightarrow$ M-theory on $R^8\times T^2 \times S^1/Z_2$, the KK modes along the nineth dimensions in the M-theory can be interpreted as the KK modes along the eighth dimension in the heterotic theory, while membrane wraping modes are winding modes of the heterotic string.

\section{Conclusions}
The searches for KK modes at the Large Hadron Collider is relevant not only for experimentators, but also for theorists, since they are related to the existence of a space of extra dimensions. Such space for the F-theory is an eightdimensional space, a special case of which can be the Calabi-Yau fourfold. It was of interest to shed light on the relationship of the Calabi-Yau fourfold structure to the appearance of a mass of KK-partners. On the basis of the duality between the heterotic string and the F-theory, as well as between the heterotic string and the M-theory, the article presents formulas for the masses of the KK modes for the K3-compactified F-theory.

\end{document}